# The Product Arbitrariness of Generalised Functions and its Role in Quantum Field Theory

Luca Nanni
luca.nanni@edu.unife.it

## Abstract
In this study, the problem of the product arbitrariness of generalised functions in the framework of Schwartz distribution is addressed. This arbitrariness is responsible for the problem of infinities encountered in quantum field theory when higher-order corrections are considered. The methods of Güttinger-König and Hörmander for getting rid of this arbitrariness are investigated.

## 1. Introduction

The problem of infinities has long been known in physics, well before the introduction of quantum theory. Just think of the energy of an electric field generated by a charged sphere whose radius tends to be zero or the mass density of a point-like particle [1–2]. In the 19th century, while studying the analytical theory of heat, Fourier had come across problems of infinity and was able to solve them by introducing a new function, known as the Delta impulse [3]. Similar problems were encountered by Gustav Kirchhoff during the formulation of the theory of light rays [4] and by Hermann von Helmholtz studying the wave equation [5]. However, it was with the emergence of the quantum field theory (QFT) that the problem of infinities became one of the main research topics, attracting not only the attention of physicists but also mathematicians [6–8]. Generally, the problem of infinities arises whenever one tries to refine the calculation of a physical theory, which, at first approximation, seems to offer promising results. Feynman, Schwinger and Tomonaga had to contend with this difficulty while formulating the theory of quantum electrodynamics [9–11]. The perturbation theory that they were working on led to divergent formal series that did not have a univocal mathematical meaning [12]. Such divergences arose because the coefficients of the formal series contained products of generalised functions, such as $\theta(x)x^{-1}$ or $\delta(x)x^{-1}$, where $\theta(x)$ and $\delta(x)$ denote the Heaviside and Dirac functions, respectively, whose values were not well defined. In quantum electrodynamics, the Feynman propagator itself is a distribution, and in the interactions between fields, some terms of the Lagrangian contain their product. The problem of divergences was solved with the renormalisation process [13–14], which consists of a series of physico-mathematical techniques that are applicable until today in the framework of statistical mechanics and fractal theory [15].

The problem of the arbitrariness of the product between generalised functions remains an intriguing research topic in the field of mathematical physics [16]. In 1954, Schwartz published a paper through which the impossibility of the multiplication of generalised functions was proved [17]. This result seemed to compromise the possibility of solving the problem of infinities, which most theoretical physicists were working on in the 1950s. However, rapid progress on the theory of distributions [18], along with Colombeau's innovative work concerning the formulation of a new theory of distributions [19], led Schwartz to reconsider his previous conclusion. In 1983, Schwartz presented to the French Academy of Sciences the note entitled *A General Multiplication of Distributions* in which the necessary conditions to define a multiplication between generalised functions were revisited [20]. In the same year,

Colombeau published an article on the multiplication of distributions, which, to date, is considered to be the piece of work that solved this mathematical problem [21].

In this study, two different approaches for the definition of a product between generalised functions that satisfy both the distributive property and the Leibnitz rule are revisited. The first approach is that of Güttinger-König [22–23], which is based on the Hahn-Banach extension theorem [24], and the second is that of Hörmander [25] via Fourier transform. These approaches do not require a reformulation of the distribution theory and are developed within the space of Schwartz distributions by exploiting known theorems of analysis. On the other hand, the Colombeau approach to distributions' multiplication requires the formulation of a new differential algebra that contains the space of Schwartz's distributions but is limited to preserving the product of smooth functions instead of bounded continuous functions. Therefore, the Güttinger-König and Hörmander approaches are complementary to that of Colombeau and are equally fundamental to solving the problems of infinity that affect the QFT.

The manuscript is organised as follows. In Section 2, the basic concepts of space of fundamental functions (or test functions) and space of Schwartz distributions are introduced. In Section 3, some of the generalised functions relevant to the QFT are discussed. In Section 4, the problem of the arbitrariness of the product of distributions is introduced and explained using a few examples. In Section 5, the Güttinger-König approach is investigated in detail. Finally, in Section 6, the Hörmander approach is introduced, and its simplicity and computing power are compared with those of the Güttinger-König approach.

## 2. Preliminaries on $\mathcal{D}$ and $\mathcal{D}'$ Spaces

In this section, the space of fundamental functions, also called space of test functions, and the space of Schwartz generalised functions are defined, and their properties and physical meanings are discussed to simplify the reading of the following sections. For a rigorous discussion on generalised functions and their operations, refer to the monumental work of Gelfand and Shilov [18].

**Definition 1**: The space of fundamental functions or test functions, denoted by $\mathcal{D}$, is the set of infinitely differentiable complex-valued functions $\varphi(x)$ on a non-empty subset $U \subset \mathbb{R}^n$ that has compact support.

The support of function φ(x), denoted by $supp(\varphi)$, is the set of point for which $\varphi(x) \neq 0$.

**Definition 2**: A sequence $\{\varphi_k\}$, with $\varphi_k \in \mathcal{D}$, converges to $\varphi(x) \in \mathcal{D}$ if $\exists R > 0$ such that $supp(\varphi_k) \subset U_R$, where $U_R$ is a ball of radius $R$ included in $\mathbb{R}^n$.

It follows that as $x \to \pm\infty$, the function $\varphi(x) \to 0$. This behaviour is fundamental in physics, although there are some examples for which this behaviour does not hold. In fact, there are partial differential equations that do not admit regular solutions with continuous derivatives. If one admits a non-differentiable function as the solution of a well-posed problem, this solution is called a weak solution that has a weak derivative defined in the Sobolev space [26]. In the measurement processes, the test functions represent the instruments that interact with the physical system being studied. These processes are local, that is, the physical quantity measured is the average value over the range of actions of the measuring instrument. Beyond this range, the measured value goes down rapidly to zero. The most common forms of test function used in physics are bump functions. In $\mathbb{R}$, such functions can be defined as follows:

$$\varphi_n(\text{x}) = \begin{cases} 0 \;\forall x \notin (-1,1) \\ exp\left[\frac{1}{x^{2n} - 1}\right] \forall x \in (-1,1) \end{cases}, \qquad (1)$$

where $n \in \mathbb{N}$. The trend of these functions as $n$ varies is represented in Figure 1.

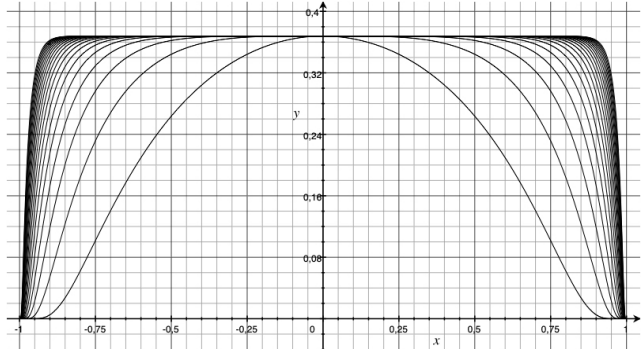
Figure 1: One-dimensional bump functions $\varphi_n(x)$ with $supp(\varphi) = (-1,1)$

Any smooth deformation of these functions is a good test function for a measurement process. As a consequence of definition 2, the following proposition holds:

**Proposition 3**: $\forall \varphi_k \in \mathcal{D}$, as $k \to \infty$, the sequence $\{\varphi_k^{(n)}\}$ tends to $\varphi^{(n)}$, where $(n)$ denotes the $n^{th}$ derivative.

By introducing constraints to the definition of the test function, subspaces of $\mathcal{D}$ that are particularly interesting for our study are obtained. For instance, let us consider a set of infinitely differentiable complex-valued functions $\varphi(x)$ that, together with their derivatives, tend to move to zero more rapidly than any power of $1/|x|$ as $x \to \infty$:

$$|x^\alpha \varphi^{(\beta)}(x)| \leq \gamma_{\alpha\beta} \quad \forall \alpha, \beta = 1, 2, \cdots, \tag{2}$$

where $\gamma_{\alpha\beta}$, is the numerical constant. This definition holds, although if $\varphi$ is an analytic function, then the inequality given by Eq. (2) becomes:

$$|z^n \varphi(z)| \leq \gamma_n exp(c|y|), \tag{3}$$

where $\gamma_n$ and $c$ are numerical constants that could depend on the choice of $\varphi$. It should be noted that if $\varphi$ is an analytic function, then the condition of convergence given by definition 1 is no longer applicable. For this class of functions, the following proposition holds:

**Proposition 4**: The sequence $\{\varphi_k(z)\}$, such that $\varphi_k(z)$ is analytic, is convergent if:

$$|z^n \varphi_k(z)| \leq \gamma_n exp(c|y|) \quad and \quad \lim_{x \to x_0} \varphi_k(x + iy) = 0 \quad \forall x_0 \in \mathbb{R}. \tag{4}$$

In other words, $\{\varphi_k(z)\}$ converges uniformly to zero in every interval of the real axis. It must be noted that in Eq. (4), the constants $\gamma_n$ and $c$ do not depend on the index $k$ of the sequence.

Let us now proceed to defining the space of a regular generalised function.

**Definition 5**: Every continuous and linear function on the space $\mathcal{D}$ of test functions is called a generalised function. The latter is said to be regular if it can be obtained from a locally integrable function $f(x)$ in $\mathbb{R}^n$ and the following integral holds:

$$(f, \varphi) = \int f(x)\varphi(x)dx \quad \forall \varphi \in \mathcal{D}. \tag{5}$$

Any other generalised function that does not satisfy the integral of Eq. (5) is considered singular. The Schwartz distribution $f(x)$ can always be obtained by the weak convergence of sequences with suitable ordinary parametric functions $\{f_\varepsilon(x)\}_{\varepsilon \in \mathbb{R}}$, which are also called mother functions [27]. The function $f_\varepsilon(x)$ can be seen as a smooth approximation of $f(x)$. The set of all functionals on the space $\mathcal{D}$ defines the space of generalised functions denoted by $\mathcal{D}'$. Therefore, the space $\mathcal{D}'$ is the dual of the space $\mathcal{D}$, that is, $\mathcal{D}' = \mathcal{D}^*$. It is clear that the dual of the subspaces of $\mathcal{D}$ discussed above are subspaces of $\mathcal{D}'$. In particular, the dual of the space of test functions satisfying the condition of Eq. (2) is the space of the tempered distributions. This subspace is particularly important in physics because it contains all the generalised functions for which the Fourier transform can be computed. Instead, the dual of the subspace of test functions satisfying the condition of Eq. (3) is defined as the space of hyperfunctions. Utilizing the latter, it is possible to formulate the QFT in terms of their Fourier transforms; thus, obtaining a formalism is useful for solving the problem of non-renormalizable interactions [28].

In measurement processes, the generalised function represents the (microscopic) physical system on which the (macroscopic) measurement instrument, represented by the test function, acts. In this context, the value obtained from the measure is the result of the action of the functional $f(x)$ on the test function $\varphi(x)$ via the formula of Eq. (5).

## 3. Main Generalised Functions of QFT

This section is devoted to the introduction of the main generalised functions encountered in QFT problems and to the discussion of their analytical properties. Let us start with the Heaviside step function, defined as:

$$\theta(x) = \begin{cases} 1 & \forall x > 0 \\ 0 & \forall x < 0 \end{cases}. \tag{6}$$

This function is not defined at $x = 0$. Using the notion of Fourier transform and the notion of Cauchy principal value, one gets its integral representation as [29]:

$$\theta(x) = -\frac{1}{2\pi i} \lim_{\varepsilon \to 0} \int_{-\infty}^{\infty} \frac{e^{-ikx}}{k + i\varepsilon} dk. \tag{7}$$

The limit inside the integral is obtained through:

$$-\frac{1}{2\pi i} \int_{-\infty}^{\infty} \lim_{\varepsilon \to 0} \frac{e^{-ikx}}{k + i\varepsilon} dk = \mathrm{P}\left(\frac{1}{x}\right) + i\pi\delta(x), \tag{8}$$

where P denotes the Cauchy principal value and $\delta(x)$ is the delta Dirac function (that will be introduced shortly). The action of $\theta(x)$ on the space of test functions is given by:

$$(\theta(x), \varphi) = \int_{0}^{\infty} \varphi(x) dx \quad \forall \varphi \in \mathcal{D}. \tag{9}$$

The integral of Eq. (9) is always convergent since test functions have limited support; beyond this support, they rapidly go down to zero.

Let us now introduce the delta Dirac function, the most widespread and known generalised function in the framework of mathematical physics. The action of $\delta(x)$ of the space $\mathcal{D}$ is given by:

$$(\delta, \varphi) = \int \delta(x)\varphi(x) dx = \varphi(0) \quad \forall \varphi \in \mathcal{D}. \tag{10}$$

The delta Dirac function is related to the Heaviside function through the derivative of the latter. Using Eq. (7), one gets:

$$\frac{d\theta(x)}{dx} = -\frac{1}{2\pi i} \lim_{\varepsilon \to 0} \int_{-\infty}^{\infty} \frac{d}{dx}\left(\frac{e^{-ikx}}{k + i\varepsilon}\right) dk = \frac{1}{2\pi} \int_{-\infty}^{\infty} e^{-ikx} = \delta(x), \tag{11}$$

where, in the last step, the integral representation of the delta Dirac function has been used. The $\delta(x)$ function can also be defined in the complex plane. In this case, the test functions are analytical. Using the Cauchy formula, one obtains:

$$(\delta(z - z_0), \varphi(z)) = \int_{-\infty}^{\infty} \delta(z - z_0)\varphi(z)dz = \frac{1}{2\pi i}\int_{\Gamma_0} (z - z_0)^{-1}\varphi(z)dz, \qquad (12)$$

where $\Gamma_0$ is a closed path. The result of Eq. (12) has the same form as that obtained from Eq. (10), namely, $(\delta(z - z_0), \varphi(z)) = \varphi(z_0)$.

Let us now introduce the class of generalised functions defined through the Cauchy principal value ansatz. The functional $\text{P}(1/x) = d\ln|x|/dx$ is a generalised function whose action on the test function $\varphi(x)$ is given by:

$$(\text{P}(1/x), \varphi) = \int_{-\infty}^{\infty} \frac{d\ln|x|}{dx}\varphi(x)dx = \lim_{\varepsilon \to 0^+}\left[\int_{-\infty}^{\varepsilon}\frac{\varphi(x)}{x}dx + \int_{\varepsilon}^{\infty}\frac{\varphi(x)}{x}dx\right]. \qquad (13)$$

Integrating Eq. (13) by part one gets:

$$(\text{P}(1/x), \varphi) = \lim_{\varepsilon \to 0^+} \int_{|x|>\varepsilon} \ln|x|\,\varphi'(x)dx, \qquad (14)$$

since $\varphi(x) \to 0$ as $x \to \pm\infty$. Eq. (14) represents the explicit action of $\text{P}(1/x)$ on the space $\mathcal{D}$. Using the same approach, the following generalised functions are also defined:

$$\begin{cases} x_+^{-1} = \theta(x)x^{-1} = \begin{cases} x^{-1} & \forall x > 0 \\ 0 & \forall x < 0 \end{cases} \\ x_-^{-1} = -\theta(-x)x^{-1} = \begin{cases} 0 & \forall x > 0 \\ x^{-1} & \forall x < 0 \end{cases} \end{cases}. \qquad (15)$$

It should be noted that both $\theta(x)$ and $x^{-1}$ are not defined at $x = 0$, and the only way to address these functions is to use the Cauchy principal value ansatz. Eq. (15) can be generalised by defining the functions $x_\pm^\lambda = \pm\theta(\pm x)x^\lambda$ with $-1 < \lambda < 1$.

## 4. Multiplication of Distributions: The Problem of Arbitrariness

Let $U \subset \mathbb{R}^n$ be an open subset of $\mathbb{R}^n$ and suppose that $\mathfrak{A}(U)$ is an associative, commutative differential algebra containing the space of distributions $\mathcal{D}'$. It is supposed that multiplication is identical to that of the ordinary functions, where the Leibnitz rule holds. In this framework, the Heaviside step function can be obtained by multiplying itself $n$ times:

$$\theta(x) = \theta^n(x) \quad \forall n \in \mathbb{N}. \qquad (16)$$

Performing the derivative of both sides of Eq. (16), one gets:

$$\frac{d\theta(x)}{dx} = \delta(x) \quad and \quad \frac{d\theta^n(x)}{dx} = n\theta^{n-1}(x)\delta(x). \qquad (17)$$

The two derivatives in Eq. (17) are different. Therefore, the equality $\theta(x) = \theta^n(x)$ does not hold. Let us consider another example:

$$\int_{-\infty}^{\infty} [\theta^2(x) - \theta(x)]\varphi(x)dx = 0 \quad \forall \varphi(x) \in \mathcal{D}. \qquad (18)$$

In the framework of the ordinary functional analysis, the integral in Eq. (18) is zero only if $[\theta^2(x) - \theta(x)] = 0$. However, one draws a completely different result if the following integral is calculated:

$$\int_{-\infty}^{\infty} [\theta^2(x) - \theta(x)]\delta(x)dx = \left[\frac{\theta^3(x)}{3} - \frac{\theta^2(x)}{2}\right]\Big|_{-\infty}^{\infty} = -\frac{1}{6}, \tag{19}$$

where $[\theta^2(x) - \theta(x)] \neq 0$. Therefore, the generalised function $[\theta^2(x) - \theta(x)]$ obtained using the product of the ordinary functions takes different values depending on the mathematical context considered, which is not possible. These examples can be generalised to any Schwartz distribution, proving that algebra $\mathfrak{A}(U)$ cannot be constructed with the product rule of ordinary functions since it suffers from arbitrariness. To better visualize the nature of this problem, let us write the $\theta(x)$ function as the weak limit of the mother function $\theta_\varepsilon(x) = 1/(1 + e^{-2x/\varepsilon})$:

$$(\theta(x), \varphi) = \lim_{\varepsilon \to 0^+} \int_{-\infty}^{\infty} \frac{1}{1 + e^{-2x/\varepsilon}} \varphi(x)dx \quad \forall \varphi(x) \in \mathcal{D}. \tag{20}$$

Using the smooth approximation $\theta_\varepsilon(x)$ of $\theta(x)$, we can also construct the following weak limit:

$$(\theta^2(x), \varphi) = \lim_{\varepsilon \to 0^+} \int_{-\infty}^{\infty} \left(\frac{1}{1 + e^{-2x/\varepsilon}}\right)^2 \varphi(x)dx \quad \forall \varphi(x) \in \mathcal{D}. \tag{21}$$

It is evident that as $\varepsilon \to 0^+$, both integrals of Eq. (20) and Eq. (21) converge to the same value, but the convergence rates are different. Therefore, the behaviour of the mother function $\theta_\varepsilon(x)$ around the point $x = 0$ is different with respect to the mother function $\theta_\varepsilon^2(x)$ and, consequently, also $\theta(x) \neq \theta^2(x)$.

To conclude this section, the following definition is given:

**Definition 6**: Two generalised functions $f(x)$ and $g(x)$, obtained by the weak convergence of the sequences $\{f_\varepsilon(x)\}$ and $\{g_\varepsilon(x)\}$, are weakly equal if the following limit holds [22]:

$$\lim_{\varepsilon \to 0} \int_{-\infty}^{\infty} [f_\varepsilon(x) - g_\varepsilon(x)]\varphi(x)dx \quad \forall \varphi(x) \in \mathcal{D}. \tag{22}$$

This definition will be particularly useful in the next section, where some of the generalised functions can be approximated by others, provided that the respective mother functions converge at a similar rate.

## 5. The Güttinger-König Method

This study aims to define a product between generalised functions that satisfies the distributive property and the Leibnitz rule without leaving any room for arbitrariness. We do not require the product to be commutative, as it is not a necessary condition to construct a well-posed algebra $\mathfrak{A}(U)$ of generalised functions. It is understood that multiplication is performed between generalised functions having the same singular point. The first that will be examined is the Güttinger-König approach, which allows the obtainment of a non-commutative algebra of the Schwartz distributions [23]. This approach is based on the Hahn-Banach extension theorem, which states the following:

**Theorem 7**: Let $\mathfrak{S}$ be a real or complex normed linear space and $A \subset \mathfrak{S}$ a subspace. Then, let $f(x)$ be a bounded linear functional, such that $f(x) \in A^*$, where $A^*$ is the dual of the subspace $A \subset \mathfrak{S}$. Then, there exists an extended functional $\tilde{f}(x) \in \mathfrak{S}^*$, where $\mathfrak{S}^*$ is the dual of space $\mathfrak{S}$, such that $\tilde{f}(x) = f(x) \ \forall x \in A$ and $\|\tilde{f}(x)\|_{\mathfrak{S}^*} = \|f(x)\|_{A^*}$.

It is important to note that to define a product rule that does not distort the vector lengths, the norm of the extended functional $\tilde{f}(x)$ is equal to the norm of the initial functional $f(x)$. This is the basic condition for eliminating the arbitrariness of the product discussed in the previous section.

Let $f(x) \in \mathcal{D}'$ be a Schwartz distribution and $\alpha(x) \in C^\infty$ a multiplier, that is, a smooth continuous infinitely differentiable function. Then, it is possible to construct the new generalised function $h(x) = \alpha(x)f(x)$ by the following product:

$$\big(\alpha(x)f(x), \varphi(x)\big) = \big(f(x), \alpha(x)\varphi(x)\big) \quad \forall \varphi(x) \in \mathcal{D}, \tag{23}$$

which is consistent with the definition given by Eq. (5). The product defined by Eq. (23) is associative, that is:

$$\big([\alpha(x)\beta(x)]f(x), \varphi(x)\big) = \big(\alpha(x)f(x), \beta(x)\varphi(x)\big) = \big(f(x), \alpha(x)\beta(x)\varphi(x)\big). \tag{24}$$

Moreover, it satisfies the Leibnitz rule as follows:

$$\begin{aligned}\big([\alpha(x)f(x)]', \varphi(x)\big) &= \big([\alpha(x)f'(x) + \alpha'(x)f(x)], \varphi(x)\big) \\ &= \big(f(x), \alpha'(x)\varphi(x)\big) + \big(f'(x), \alpha(x)\varphi(x)\big).\end{aligned} \tag{25}$$

In Eq. (23), both $\alpha(x)$ and $\beta(x)$ are multipliers. The multiplication rule given by Eq. (23) is a good starting point for defining the product between distributions that satisfy the conditions of Eq. (24) and Eq. (25). Therefore, in this framework, the algebra $\mathfrak{A}(U)$ is a sub-algebra of the distributions. From now on, the product of generalised functions will be denoted by the symbol " • ". Let us start with the product between the Heaviside and delta Dirac functions. Supposing $\theta(x)$ behaves like a multiplier and using the product rule of Eq. (23), one gets:

$$\big(\theta(x) \bullet \delta(x), \varphi(x)\big) = \big(\delta(x), \theta(x)\varphi(x)\big) = \theta(0)\big(\delta(x), \varphi(x)\big) = \theta(0)\varphi(0), \tag{26}$$

which, as expected, is meaningful given that $\theta(x)$ is not defined at $x = 0$. However, Eq. (26) returns meaningful if $\varphi(0) = 0 \, \forall \varphi(x) \in U \subset \mathcal{D}$. This means that the generalised function $h(x) = \theta(x) \bullet \delta(x)$ is known for $U \subset \mathcal{D}$. Assuming that $h(x)$ is a linear continuous functional, applying the Hahn-Banach extension theorem is possible to construct its extension $\tilde{h}(x) \in \mathcal{D}'$ over the whole space $\mathcal{D}$ if its value is known on any $\psi_0 \in \mathcal{D}/U$. In other words, we can say that $h(x)$ can be extended over $\mathcal{D}$ if an arbitrary value of $h(x)$ is given on $\psi_0$. This is the key step of the Güttinger-König method. Based on what has just been argued, the test function $\psi(x) \in \mathcal{D}$ can be written as:

$$\psi(x) = \psi(0)\psi_0(x) + \varphi(x), \tag{27}$$

where $\psi_0(x) \in \mathcal{D}/U$ while $\varphi(x) \in \mathcal{D}$. Applying the functional $\tilde{h}(x)$ to the test function of Eq. (27), one gets:

$$\big(\tilde{h}(x), \psi(x)\big) = \psi(0)\big(\tilde{h}(x), \psi_0(x)\big) + \big(h(x), \varphi(x)\big). \tag{28}$$

In writing Eq. (28), it must be recalled that $\tilde{h}(x) = h(x)$ whenever $x$ belongs to the $supp(\varphi)$. Setting $\psi_0(0) = 1$ and considering that $\big(\tilde{h}(x), \psi_0(x)\big)$ is arbitrary, say equal to a constant $\kappa$, it follows that:

$$\tilde{h}(x) = \theta(x) \bullet \delta(x) = \kappa\delta(x) + 0 = \kappa\delta(x), \tag{29}$$

which holds $\forall x \in supp(\psi)$. Following the same reasoning, it is easy to prove that:

$$\tilde{h}(x) = \theta(x) \bullet x^{-1} = -\delta'(x) + \kappa\delta(x), \tag{30}$$

where the relation $x\delta(x) = -\delta'(x)$ has been used [27].

This result can be generalised as follows: let $\psi(x) \in \mathcal{D}$, $\varphi(x) \in U \subset \mathcal{D}$ and $\psi_n(x) \in \mathcal{D}/U$ be test functions. The $\mu - th$ derivative of $\psi_n(x)$ must be such that $\psi_n^{(\mu)}(0) = \delta_{n\mu}$, where $\delta_{n\mu}$ is the Kronecker delta. Then, the test function $\psi(x)$ can be written as:

$$\psi(x) = \sum_{n=0}^{2m} \psi^{(n)}(0)\psi_n(x) + \varphi(x). \tag{31}$$

Let $f(x)$ and $g(x)$ be two generalised functions having the same common singular points $x_\mu$, by which the distribution $h(x) = f(x) \bullet g(x)$ is constructed. Supposing that $h(x)$ is known on the subspace $U \subset \mathcal{D}$, it is possible to obtain the extended functional $\tilde{h}(x) = [f(x) \bullet g(x)]_{ex.}$ as:

$$\tilde{h}(x) = h(x) + \sum_{\mu,\nu} a_{\mu\nu} \delta^{(\nu)}(x - x_\mu), \tag{32}$$

where $a_{\mu\nu}$ are arbitrary coefficients. Applying the extended generalised function of Eq. (32) to the test function of Eq. (31), and considering for simplicity $x_\mu = 0$, $\nu = 0$, $n = 0$ and $\psi^{(0)}(0) = 1$, one gets:

$$\left(\tilde{h}(x), \psi(x)\right) = \left(h(x) + a_{00}\delta(x), \psi_0(x) + \varphi(x)\right), \tag{33}$$

which can be further developed as:

$$\left(\tilde{h}(x), \psi(x)\right) = \left(h(x), \psi_0(x)\right) + \left(h(x), \varphi(x)\right) + \left(a_{00}\delta(x), \psi_0(x)\right) + \left(a_{00}\delta(x), \varphi(x)\right). \tag{34}$$

Considering that the value of $\left(h(x), \psi_0(x)\right)$ is arbitrary, say $\kappa$, and assuming that the known value of $\left(h(x), \varphi(x)\right)$ is zero $\forall x \in supp(\varphi)$, Eq. (34) becomes:

$$\left(\tilde{h}(x), \psi(x)\right) = \kappa + a_{00}\psi_0(0). \tag{35}$$

It can be immediately verified that Eq. (35) is consistent with the product in Eq. (30). Eq. (35) is the generalization of the Güttinger-König method which, to the best of our knowledge, does not appear in any other published work on the subject, not even in the original ones by König [22]. This equation allows a simple and immediate way to calculate the product of generalized functions that verify the conditions required by Theorem 7. The arbitrariness of the choice of constants $\kappa$ and $a_{\mu\nu}$, as well as the hypothesis of considering $\left(h(x), \varphi(x)\right)$ zero or not zero, fails when the problem is contextualised as per the given physical scenario. In this case, it will be the boundary conditions of the problem that shall constrain the arbitrary coefficients to assume only certain values. A rigorous product theory that verifies the Leibnitz rule, even if it is not associative or commutative, has been obtained. Notably, the associative properties fail because given three generalised functions, namely, $f(x)$, $g(x)$ and $u(x)$, if the products $f(x) \bullet g(x)$ and $g(x) \bullet u(x)$ yield the generalised functions $h(x)$ and $q(x)$, respectively, the products $h(x) \bullet u(x)$ could be different from the product $f(x) \bullet q(x)$. To be convinced of this is sufficient to iterate the product rule given by Eq. (32). Through the definition given by Eq. (33), König constructed a non-associative and non-commutative algebra containing the Schwartz distributions [23].

The Güttinger-König product rule has been obtained strictly within the Schwartz distribution space without the need to introduce subsidiary conditions. Furthermore, this rule is applicable to any Schwartz function, unlike that of Colombeau, which preserves the product only for smooth functions and not for bounded continuous functions, and that of Hörmander, which is applicable only to distributions that admit Fourier transform. The main difficulty of the Güttinger-König method is determining a priori the function $h(x)$ locally defined on the $supp(\varphi)$ $\forall \varphi \in U \subset \mathcal{D}$. This difficulty can be overcome by using the weak equality principle, defined by Eq. (22), and considering that $supp(\varphi)$ contains the singular point of the two distributions to be multiplied. The key point is to study the convergence rates of the product of the mother functions associated with the distributions to be multiplied concerning the convergence rate of an adequate mother function associated with the $h(x)$ the function we are looking for. To clarify this concept, let us compute the product $\delta(x) \bullet \delta(x)$. We need to find the explicit form of $h(x)$ in a set $I(0,r)$ with $r \ll 1$. This set contains the singular point $x = 0$ and represents the $supp(\varphi)$, where the values $(h(x), \varphi(x))$ must be known. A suitable mother function for the delta Dirac function is the Gaussian function:

$$\delta_\varepsilon(x) = \frac{1}{\sqrt{2\pi\varepsilon}} exp\left(-\frac{x^2}{2\varepsilon}\right). \tag{36}$$

Therefore, the mother function associated with the product $\delta(x) \bullet \delta(x)$ is $[\delta_\varepsilon(x)]^2$ is:

$$[\delta_\varepsilon(x)]^2 = \frac{1}{2\pi\varepsilon} exp\left(-\frac{x^2}{\varepsilon}\right). \tag{37}$$

Let us now calculate the derivative of $\delta_\varepsilon(x)$:

$$\delta'_\varepsilon(x) = -\frac{x}{\varepsilon\sqrt{2\pi\varepsilon}} exp\left(-\frac{x^2}{2\varepsilon}\right), \tag{38}$$

from which one also gets:

$$-\frac{\delta'_\varepsilon(x)}{x} = \frac{1}{\varepsilon\sqrt{2\pi\varepsilon}} exp\left(-\frac{x^2}{2\varepsilon}\right). \tag{39}$$

Once set $x_0 \in I(0,r)$ is given, the two parametric functions of Eq. (37) and Eq. (38) converge with the same rate as $\varepsilon \to 0$:

$$\lim_{\varepsilon \to 0} \int_{-\infty}^{\infty} \left[\frac{1}{2\pi\varepsilon} exp\left(-\frac{x_0^2}{\varepsilon}\right) - \frac{x_0}{\varepsilon\sqrt{2\pi\varepsilon}} exp\left(-\frac{x_0^2}{2\varepsilon}\right)\right] \varphi(x) dx = 0 \quad \forall \varphi \in U \subset \mathcal{D}. \tag{40}$$

This result is shown in Figure 2, where the trends of $[\delta_\varepsilon(x_0)]^2$ (soft grey dotted line), $\delta_\varepsilon(x_0)$ (soft blue dotted line) and $\{[\delta_\varepsilon(x_0)]^2 - \delta_\varepsilon(x_0)\}$ (blue dotted line) are plotted vs. $\varepsilon$ for $x_0^2 = 0{,}1$:

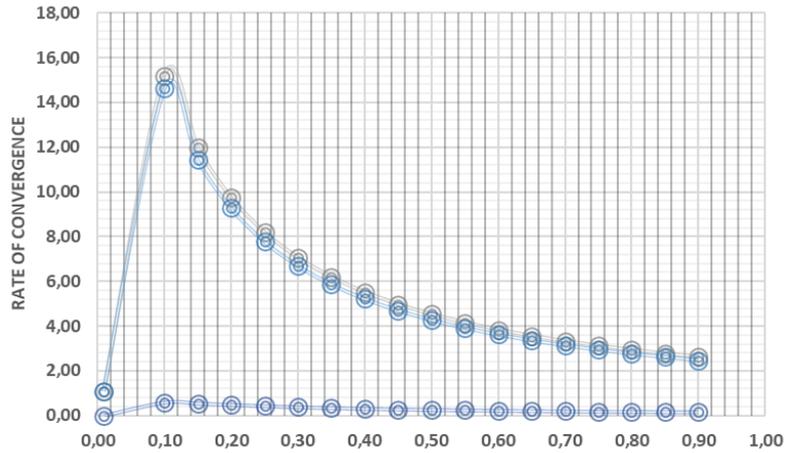

Figure 2: $[\delta_\varepsilon(x_0)]^2$ (soft grey dotted line), $-\delta'_\varepsilon(x_0)/x_0$ (soft blue dotted line) and $\{[\delta_\varepsilon(x_0)]^2 - |-\delta'_\varepsilon(x_0)/x_0|\}$ (blue dotted line) vs. $\varepsilon$

Since $-\delta'_\varepsilon(x)/x = \delta_\varepsilon(x)$, one concludes that $h(x) \approx \delta(x) \ \forall x \in I(0, r)$. Therefore, using Eq. (32), we can expand $h(x)$ obtaining the following:

$$\delta(x) \bullet \delta(x) = (1 + a_{00})\delta(x) + a_{01}\delta^{(1)}(x), \qquad (41)$$

This brings to mind the Taylor series development. It is evident that Eq. (41) approximates the product $\delta(x) \bullet \delta(x)$ since $\delta(x)$ is not equal but weakly equal to $h(x)$ in the set $I(0, r)$. However, in physics, this approximation is adequate. This (approximated) variant of the Güttinger-König method allows us to calculate the product $\delta(x) \bullet \delta(x)$, which is otherwise not obtainable through other approaches.

## 6. The Hörmander Method

The Hörmander approach to the distribution multiplication [25] is simpler and more immediate than that of Güttinger-König. Furthermore, it provides a product that is associative and commutative; thus, it is more akin to the Colombeau method. However, being based on the Fourier transform, it can be applied only to the tempered generalised functions that have been introduced in Section 2. In this sense, the Hörmander method is more restrictive than the Güttinger-König one. Let $f(x)$ and $g(x)$ be two tempered generalised functions; then their product is defined as:

$$f(x) \bullet g(x) = \frac{1}{(2\pi)^n} \mathfrak{F}^{-1}\{\mathfrak{F}[f(x)] * \mathfrak{F}[g(x)]\} \quad \forall x \in \mathbb{R}^n, \qquad (42)$$

where $\mathfrak{F}$ and $\mathfrak{F}^{-1}$ denote the Fourier transform and antitransform respectively, whereas " $*$ " denotes the convolution product. The product defined by Eq. (42) exists only if the convolution integral is convergent. For clarity, the explicit definitions of Fourier transform and the convolution product of generalised functions in $\mathbb{R}$ are [30–31]:

$$\mathfrak{F}[f(x)] = \int f(x)e^{ikx}dx \ | \ \big(\mathfrak{F}[f(x)], \varphi(x)\big) = \mathfrak{F}[\varphi(x)] \quad \varphi(x) \in \mathcal{D}, \qquad (43)$$

and

$$f(x) * g(x) = \int f(y)g(x - y)dy \quad f(x), g(x) \in \mathcal{D}'. \qquad (44)$$

It should be noted that the product $f(y)g(x - y)$ is the direct product between generalised functions, as defined by Eq. (23), and should not be confused with the multiplication rule between generalised functions, which is the subject of this study. The convolution product is defined for locally integrable functions; therefore, both $f(y)$ and $g(x - y)$ can be seen as multipliers [27]. Since the convolution of generalised functions is associative and commutative [31], the Hörmander multiplication rule also satisfies these properties. It also follows that the algebra of the tempered distributions endowed with the Hörmander product is associative and commutative. Moreover, the Leibnitz rule does hold for the product defined by Eq. (42). To prove this, the following properties of the convolution product and the Fourier transform must be considered: $(f(x) * g(x))' = f(x) * g'(x)$ and $f(x) * g(x) = \mathfrak{F}^{-1}\{\mathfrak{F}[f]\mathfrak{F}[g]\}$, where the product between $\mathfrak{F}[f]$ and $\mathfrak{F}[g]$ is intended as an ordinary product between continuous functions.

Among the tempered generalised functions, there is a class which, except for a limited number of isolated points, are boundary values of analytic functions of step functions. For these functions, the Hörmander product is:

$$\big(f(x) \bullet g(x), \varphi(x)\big) = Res_{z=0}\left[\frac{1}{z}\int f(x)g(x)\left(\frac{a}{x}\right)^z \varphi(x)dx\right], \qquad (45)$$

where $z \in \mathbb{C}$ and $a$ are arbitrary scaling parameters having the same dimension of $x$. For this class of generalised functions, the Fourier transform is given by:

$$\mathfrak{F}[f(x)] = Res_{z=0}\left[\frac{1}{z}\int_0^\infty f(x)\left(\frac{a}{x}\right)^z e^{-ikx}dx\right]. \quad (46)$$

The only arbitrariness of the product given by Eq. (45) is represented by the choice of the constant $a$, which must be made contextually to the physical problem being investigated.

Let us consider the product $\delta(x) \bullet \delta(x)$ computed in the framework of the Hörmander product rule. Applying the formula of Eq. (42), one obtains:

$$\delta(x) \bullet \delta(x) = \frac{1}{(2\pi)}\mathfrak{F}^{-1}\{\mathfrak{F}[\delta(x)] * \mathfrak{F}[\delta(x)]\}. \quad (47)$$

Since the Fourier transform of $\delta(x)$ is $\mathfrak{F}[\delta(x)] = 1$, the convolution product in Eq. (47) becomes:

$$\int \hat{\delta}(k)\hat{\delta}(k-q)\,dq = \int dq \quad (48)$$

where $\hat{\delta}$ denotes the Fourier transform of the Dirac function. The integral of Eq. (48) is divergent. Therefore, in the framework of the Hörmander method, as expected, the product $\delta(x) \bullet \delta(x)$ does not exist. More generally, any $n-th$ power of the Dirac function is never computable, except with the Güttinger-König method. Things would change if we multiply the derivatives of the Dirac functions, for which the Fourier transform is given by $\mathfrak{F}[\delta^{(n)}(x)] = (-ik)^n\theta(k)$. In fact:

$$\delta^{(n)}(x) \bullet \delta^{(m)}(x) = \frac{n!\,m!}{2\pi i(n+m+1)!}\delta^{(n+m+1)}(x), \quad (49)$$

provided that $(n+m+1) \geq 2$. Using Eq. (49), one can easily compute the product $\delta(x) \bullet \delta'(x) = \delta^{(2)}(x)/4\pi i$. This result suggests that the product $\delta(x) \bullet \delta(x)$, which is not calculable by the formula of Eq. (42), is somehow proportional to the first derivative of the Dirac function. This (speculative) assumption is confirmed by Eq. (41), which provides the (approximate) product $\delta(x) \bullet \delta(x)$. Using the $\delta^2(x)$ mother function given by Eq. (37) and the $\delta'(x)$ mother function given by Eq. (38), it is observed that their rate of convergence is comparable within the set $I(0,r)$. This behaviour is clearly shown in Figure 3, where the trends of $\delta^2_\varepsilon(x_0)$ (blue dotted line), $\delta'_\varepsilon(x_0)$ (soft blue dotted line) and $\{\delta^2_\varepsilon(x_0) - |\delta'_\varepsilon(x_0)|\}$ (soft grey dotted line) are plotted vs. $\varepsilon$ for $x_0^2 = 0{,}1$:

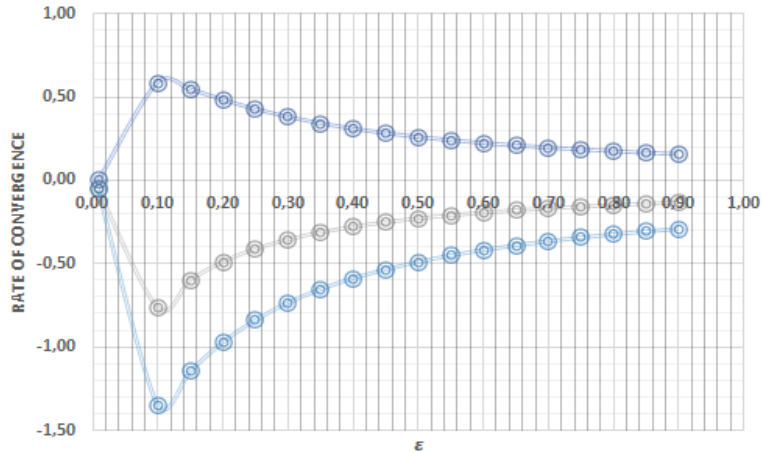

Figure 3: $\delta'_\varepsilon(x_0)$ (soft blue dotted line), $\delta_\varepsilon^2(x_0)$ (blue dotted line)

Let us now consider the product $\theta(x) \bullet x^{-1}$, which has also been calculated in the previous section with the Güttinger-König method. It can be computed by applying the formula of Eq. (45):

$$\theta(x) \bullet x^{-1} = Res_{z=0}\left[\frac{1}{z}\int \theta(x)x^{-1}\left(\frac{a}{x}\right)^z dx\right]. \tag{50}$$

Using Eq. (15), one gets:

$$\theta(x) \bullet x^{-1} = Res_{z=0}\left[\frac{1}{z}\int x^{-1}\left(\frac{a}{x}\right)^z dx\right] = \mathrm{P}(1/x). \tag{51}$$

The result of Eq. (51) can be generalised for product $\theta(\pm x) \bullet x^\lambda$, which yields the generalised function $x_\pm^\lambda$ with $-1 < \lambda < 1$ introduced in Section 3.

To conclude, let us consider another important example that is systematically encountered in the study of the interaction processes between quantum fields, represented by the product $(x + i0^+)^{-1} \bullet (x + i0^+)^{-1}$. To handle this product, the distribution $(x + i0^+)^{-1}$ must first be transformed into a more comprehensible form. For such purpose, let us consider the parametric function $(x + i\varepsilon)^{-1}$ with $\varepsilon \to 0^+$, which can be represented in the typical form $a(x) + ib(x)$. Using the algebra of complex numbers:

$$(x + i\varepsilon)^{-1} = \frac{x}{x^2 + \varepsilon^2} - i\frac{\varepsilon}{x^2 + \varepsilon^2}. \tag{52}$$

The function of Eq. (52) represents the mother function whose weak limit is the propagator $(x + i0^+)^{-1}$. This limit can be separated into two parts:

$$((x + i0^+)^{-1}, \varphi(x)) = \lim_{\varepsilon \to 0^+}\int \frac{x}{x^2 + \varepsilon^2}\varphi(x)dx - i\lim_{\varepsilon \to 0^+}\int \frac{\varepsilon}{x^2 + \varepsilon^2}\varphi(x)dx. \tag{53}$$

The first term on the right-end-side corresponds to the generalised function $\mathrm{P}(1/x)$, whereas the second term represents the distribution $-i\pi\delta(x)$ [32]. Therefore, $(x + i0^+)^{-1} = \mathrm{P}(1/x) + i\pi\delta(x)$, which is similar to Eq. (7) introduced in Section 3. The Fourier transforms (we are in the complex plane and the Fourier integral must be calculated on a contour path which does not enclose the singularities) of $\mathrm{P}(1/x)$ and $\delta(x)$ are both known [30] and lead to the obtainment of the Fourier transform $\mathfrak{F}[(x + i0^+)^{-1}] = -2\pi i\theta(-k)$. Applying the formula of Eq. (42), one obtains:

$$(x + i0^+)^{-1} \bullet (x + i0^+)^{-1} = -2\pi\mathfrak{F}^{-1}\{\theta(-k) * \theta(-k)\}. \tag{54}$$

Using the integral representation of $\theta(k)$ given by Eq. (7), it can be seen immediately that as $\varepsilon \to 0^+$, then $\mathfrak{F}^{-1}\{\theta(-k) * \theta(-k)\} = (x + i0^+)^{-2}$. However, $(x + i0^+)^{-2}$ is nothing but the first derivative of the propagator $(x + i0^+)^{-1}$. Therefore, the final result of the product $(x + i0^+)^{-1} \bullet (x + i0^+)^{-1}$ is:

$$(x + i0^+)^{-1} \bullet (x + i0^+)^{-1} = -2\pi\frac{d(x + i0^+)^{-1}}{dx} = -2\pi[\mathrm{P}'(1/x) + i\pi\delta'(x)], \tag{55}$$

where $\mathrm{P}'(1/x) = -\mathrm{P}(1/x^2)$. Following the same approach can prove that the product $(x + i0^+)^{-1} \bullet (x - i0^+)^{-1}$ does not exist. In computing the convolution integral, a divergent term $\int dq$ is encountered. Both of these examples prove the power and simplicity of Hörmander's method, which is more advantageous than other approaches.

## 7. Conclusions

Solving the product between generalised functions is a fascinating task, especially in physics and related disciplines, where the boundary conditions imposed by the system being investigated can facilitate (through appropriate approximations) or further complicate the calculation. This is not a case that new papers and books still publish, claiming developments or reinterpretations of the theories of distributions developed in the last century [33–39]. The most promising multiplication rule between distributions is defined in the Colombeau algebra, which is commutative, associative and satisfies the Leibnitz rule [21]. The main idea of Colombeau is to associate to each distribution a set of smooth parametrised functions obtained via convolution with a mollifier. The Colombeau algebra is a factor space, and each element is an equivalence class of smooth functions satisfying some moderateness estimates with respect to the parameter $0 < \varepsilon \leq 1$. However, there are other methods that can be used to define the product between Schwartz distributions which, even if they do not always satisfy the distributive property or do not admit direct solutions, nevertheless allow us to give meaning to multiplications, such as $\delta(x) \bullet \delta(x)$, which are otherwise not calculable or difficult to compute. The first of these methods is that of Güttinger-König, which has the peculiarity of being laborious but very effective in problems of physics or related sciences in which the product between distributions is a puzzle. In this framework, it has been proven that the use of the mother functions associated with the distribution to be multiplied and the study of their convergence near the singularity lead to an easy application of the Hahn-Banach extension theorem on which the Güttinger-König method is formulated. Using this technique, which is the main novelty of this paper, it becomes possible to obtain approximate results whatever the Schwartz distributions are. Furthermore, in the present manuscript, the formulas given by Eq. (29) and Eq. (30) have been generalized, proposing a development of the test function $\psi(x)$ as a finite sum of continuous and bounded functions that extend the test function $\varphi(x)$ beyond its domain of definition. The algebra constructed using the Güttinger-König multiplication rule is not associative nor commutative, but it satisfies the Leibnitz rule.

The second method is that of Hörmander, which is based on the use of the Fourier transforms of tempered distributions. Notably, even if it is more limited compared with the Güttinger-König approach, it is extremely simple and powerful (in the literature, all the Fourier transforms of the main distributions of physics-mathematics are reported), not to mention applicable to a wide range of QFT problems. In particular, in the present manuscript, we have explicitly calculated in an easy way the products of propagators $(x + i0^+)^{-1}$ and $(x + i0^+)^{-1}$, which are systematically encountered in the theoretical study of the interaction between quantum fields, which are otherwise not computable with the Güttinger-König method and in any case not easy to solve in the framework of Colombeau algebra. It fails whenever the convolution integral of the Fourier transforms diverges. However, in these cases, it is always possible to solve the problem by shifting to the Güttinger-König method. In this sense, the two methods are complementary. The algebra constructed by the Hörmander multiplication rule is commutative, associative and satisfies the Leibnitz rule and, as such, is a sub-algebra of the Colombeau algebra.

This article provides a bridge between the Colombeau theory of multiplication of generalised functions and the *old* theory of Schwartz distributions. It should be recalled that the Colombeau algebra contains the space of Schwartz distributions, but the multiplication rule that defines it preserves the product of smooth functions and not those of bound functions. This is the reason why the multiplication rules of Güttinger-König and Hörmander cannot be placed in the background, at least in the field of quantum field physics.